\documentclass[12pt,preprint]{aastex}

\shorttitle{Be Star Disks in the SMC}
\shortauthors{Ahmed \& Sigut}

\begin{document}

\setcounter{page}{2}

%\title{Be Star Disks at Low Metallicity}
\title{The Temperature Structure of Be Star Disks in the Small Magellanic Cloud}

\author{A.\ Ahmed and T.\ A.\ A.\ Sigut}
\affil{Department of Physics and Astronomy, The University of 
Western Ontario \\ London, Ontario, CANADA N6A 3K7 }
\email{ahamaz5@uwo.ca \\asigut@uwo.ca}

\slugcomment{Accepted in ApJ October 3, 2011}

\begin{abstract}

The temperature structure of Be star circumstellar disks at the
sub-solar metallicity appropriate to the Small Magellanic Cloud (SMC)
is investigated.  It is found that for central stars of the same
spectral type, Be star disks in the SMC are systematically hotter by
several thousand degrees compared to Milky Way (MW) disks with the same
density structure.  For early spectral types (B0e -- B3e), this results
in systematically smaller H$\alpha$ equivalent widths for Be stars in
the SMC. The implication of this result on Be star frequency comparisons
between MW and SMC clusters is shown to be a  5 -- 10\% lowering of the
detection efficiency of Be stars in SMC clusters.  These calculations are
also compared to the known H$\alpha$ equivalent width distributions in the
MW and SMC. For the MW, reasonable agreement is found; however, for
the SMC, the match is not as good and systematically larger Be star
disks may be required.

\end{abstract}


\section{Introduction}

Be stars are defined observationally as non-supergiant B stars that
have, or have had in the past, emission detected in the Balmer series
of hydrogen, most notably in H$\alpha$ \citep{slet88}. The emission
is thought to arise from a thin, equatorial disk of gas surrounding
the central star \citep{por03}. This picture is consistent with many
other properties of Be stars, such as an excess of infrared radiation
\citep{cote87}, linear continuum polarization \citep{mcl78,pok79},
and interferometric observations that directly resolve the disks
\citep{quir97,tyc05}.

In the Milky Way about 17\% of all non-supergiant B stars are Be stars,
although the fraction varies widely with spectral type, reaching a
maximum of $\sim\!34$\% at B1 \citep{zor97}.  Why some B stars, and
not others, become Be stars is currently unclear. Rapid rotation of
the central B star seems to play a key role \citep{por03}, but details
are still lacking as a definitive determination of the actual rotation
rates of the central stars is complicated by the potential effects of
gravitational darkening \citep{tow04,cra05,fre05}.

Also unclear is the exact mechanism(s) that creates Be star disks.
Keplerian rotation, now established for Be star disks \citep{hum00,oud08},
suggests viscous disks as the correct physical model \citep{lee91,por99},
but how material is feed into the inner edge of the disk is unknown
\citep{cra09,krt11}. It is possible pulsation plays a critical role
\citep{cra09}, and there is tantalizing observation evidence supporting
this view \citep{riv98,hua09}. The role of binarity in the Be phenomena
is unclear \citep{por03}, although the majority of Be stars do not
seem to be the result of binary evolution \citep{vbv97}. In this current
work, we consider only single, isolated Be stars.

Key to understanding the factors that produce the Be phenomenon are
cluster studies which allow ages to be assigned to the individual Be
stars.  Then any trends with evolutionary status can be examined. There
have been claims either that the Be phenomenon occurs primarily in the
later half of the main sequence \citep{fab00,fab05,mcs05a,mar06} or
throughout the entire main sequence \citep{kel99,mar07a}. Other studies
find Be stars both in very young clusters, $<10\;$Myr, and an enhancement
in older clusters, $20-30\;$Myr \citep{wis06,mat08}, suggesting two
effects at work: some Be stars are born as rapid rotators while others
become rapid rotators during main sequence evolution due to the internal
redistribution of angular momentum \citep{eks08}.

Be stars are also detectable in open clusters in the Large and Small
Magellanic Clouds (LMC and SMC), and this allows investigation of
the role of metallicity in the Be phenomena.  The solar metallicity
is $z=0.02$ while the metallicity of the LMC is about $0.007$ and the
SMC, $0.002$ \citep{mae99}. The trend of the fraction of Be
stars in a cluster with metallicity seems clear from many studies:
the fraction of Be stars tends to rise with decreasing metallicity
\citep{mae99,wis06,mar06,mar07a,mar10}.  Such a trend may have a natural
explanation in the dependence of mass-loss due to stellar winds on the
metallicity \citep{vin01,pen04}; a lower metallicity results in weaker
stellar winds and hence stars in low-metallicity environments rotate more
quickly, a fact well established for the SMC \citep{hun08}. Thus the more
quickly rotating B star population leads to a higher fraction of Be stars.

Nevertheless, all of these comparative studies between the MW,
LMC, and SMC proceed by counting candidate Be stars in the total
cluster population, selected either photometrically \citep[see, for
example,] []{mcs05b} or spectroscopically with very low-resolution
\citep[see][]{mar10}. Of course not all Be stars in a population can be
found this way and detection efficiencies generally point to selection
criteria of a H$\alpha$ emission equivalent width\footnote{We adopt
the convention that a positive equivalent width denotes emission.} in
excess of  $5\!-\!10\,$\AA\ as required to identify a candidate Be star
\citep{mar10}. Assumed in all of these studies is that individual Be stars
are more or less the same in the MW, LMC, and SMC so that the comparative
statistics are not affected by the detection efficiency. However, there
are indications that, at some level, this may not be true. For example,
the equivalent width distribution of H$\alpha$ differs between the MW
and SMC, with the SMC distribution peaking at higher equivalent widths
\citep{mar07b}.

In this work, we will illustrate a fundamental difference between Be star
disks in the MW and the SMC: for the same central star spectral type and
disk density distribution, disks in the SMC are much hotter than in the
MW. We will quantify this effect over the early Be star spectral types
(those for which LMC and SMC samples are reasonably complete), and we
will examine how counts of the Be star populations based on the H$\alpha$
equivalent width in different metallicity systems might be affected by
this systematic temperature difference.

\section{Calculations}

The thermal structure of the Be star disks was computed with the {\sc
bedisk} code of \cite{sig07}. This code enforces radiative equilibrium
in a photoionized circumstellar disk including the heating and cooling
processes for the 9 most abundant elements (H, He, CNO, Mg, Si, Ca, \&
Fe) over several ionization stages.  Details of the atomic models and
atomic data, as well as an overview of the {\sc bedisk} code, are given
in \citet{sig07}.

The density structure of the disk was assumed to be of the fixed form
\begin{equation} 
\rho(R,Z) =
  \rho_o \left(\frac{R_*}{R}\right)^{n} e^{-\left(\frac{Z}{H}\right)^2}
               \;,
\label{eq:rho} 
\end{equation} 
where $R$ and $Z$ are the cylindrical co-ordinates for the assumed
axisymmetric disk, and $R_*$ is the radius of the central B star.
The quantities $n$ and $\rho_o$ are adjustable parameters that fix
the density structure of the disk. The function $H$, defined below by
Eq.~\ref{eq:scale_height}, sets the vertical (or $Z$) scale height of the
disk.  This simple density model has been very successful in interpreting
a wide range of Be star observations \citep{gie07,tyc08,jon08}.  Thus the
thermal models of this work are identified first by the spectral type
of the central B star (which sets the photoionizing radiation field)
and then by the disk density parameters $\rho_o$ and $n$.

The fundamental parameters assumed for the central B stars are
given in Table~\ref{tab:bstars}. The mass and radius as a function
of spectral type, as well as the MW $T_{\rm eff}$ calibration, are
adopted from \cite{cox00}.  Note that we have assumed the same mass
and radius as a function of spectral type for the MW and SMC. Stellar
evolution calculations have shown that B stars are likely smaller in
low metallicity systems such as the SMC. For example, \cite{mae01} show
that a $20\,\rm M_{\sun}$ star (roughly a B0 star) decreases in radius
by about 20\% as the metallicity is lowered from $Z=0.02$ to $Z=0.004$.
However, we have ignored these differences to ensure both the SMC and
MW disk models have the same density structure. The spectral-type --
$T_{\rm eff}$ calibration for the SMC is discussed below.

The vertical (or $Z$) dependence of Eq.~\ref{eq:rho} contains
the function $H$ which is defined as
\begin{equation}
H=\left[\frac{2R^3}{GM_*}\frac{kT_d}{\mu m_{\rm H}}\right]^{1/2} \;.
\label{eq:scale_height}
\end{equation}
Here $M_*$ is the mass of the central B star, $\mu$ is the mean-molecular
weight of the disk gas, and $T_d$ is an assumed isothermal disk
temperature. Eqs.~\ref{eq:rho} and \ref{eq:scale_height} follow from
the assumption that the disk is in vertical hydrostatic equilibrium set
by the $Z$ component of the stellar gravitational acceleration and an
assumed isothermal disk temperature, $T_d$. {\it Note that $T_d$ is only used 
in Eq.~\ref{eq:scale_height} to
fix the vertical density structure of the disk}; the actual temperatures
in the disk, $T(R,Z)$, are found by enforcing radiative equilibrium.

\citet{sig09} consider consistent models in which the vertical structure
of the disk is found by integrating the equation of hydrostatic
equilibrium in a manner consistent with the radiative equilibrium disk
temperatures; this treatment eliminates the need for the parameter
$T_d$. However, the disk density distribution is now dependent on the
temperatures in the disk. As the desire of the current work is to compare
the temperature of MW and SMC disks of identical density structure,
we have used Eqs.~\ref{eq:rho} and \ref{eq:scale_height} with the same
$T_d$ for each spectral type (namely $0.6\,T^{\rm MW}_{\rm eff}$ where
$T^{\rm MW}_{\rm eff}$ is the Milky Way effective temperature at a given
spectral type).\footnote{Their remains a difference between the MW
and SMC density structures due to the different mean-molecular weights
of the gas, but this difference is extremely small.} We do not expect
this assumption to significantly affect our differential comparison.

\begin{deluxetable}{crrrr}
\tablewidth{0pt}
\tablecaption{Parameters adopted the central B stars.\label{tab:bstars}}
\tablehead{
\colhead{Spectral Type} & \colhead{Mass}  & \colhead{Radius} & 
\colhead{$T^{\rm MW}_{\rm eff}$} & \colhead{$T^{\rm SMC}_{\rm eff}$} \\
~ & \colhead{($M_{\sun}$)} & \colhead{($R_{\sun}$)} & \colhead{(K)} & \colhead{(K)}
}
\startdata
B0      & 17.5  & 7.4 & 30000  & 32000 \\
B0.5    & 15.4  & 6.9 & 28000  & 29600 \\
B1      & 13.2  & 6.4 & 25000  & 28000 \\
B1.5    & 11.0  & 5.9 & 23000  & 26400 \\
B2      &  9.1  & 5.3 & 21000  & 23800 \\
B3      &  7.6  & 4.8 & 19000  & 22000
\enddata
\tablecomments{The mass and radius calibrations are taken from \cite{cox00} and assumed
to be the same for the MW and SMC. The MW $T_{\rm eff}$ calibration is from \cite{cox00}
and the SMC calibration is based on \cite{tru07} (see text).}
\end{deluxetable}

The disk is assumed to be in Keplerian rotation and hence rotationally
supported in the $R$ direction. Gravitational darkening due to the
potential rapid rotation of the central B star is not included, although
this process can affect the thermal structure of the disk \citep{mcg11}.

The main energy input into the disk is assumed to be photoionizing
radiation from the central star. For this paper, the older \cite{kur93}
LTE photoionizing fluxes used by \citet{sig07} were replaced with the
newer non-LTE calculations of \cite{Hubeny07}. Two grids from these
calculations were used, the solar metallicity grid (p00) to compute
the MW models, and the $1/10$ solar abundance grid (m10) to compute the
SMC models.

The thermal disk models are thus described by three parameters:
the spectral type of the central B star (which is assumed to fix the
stellar mass, radius and effective temperature), and the two parameters
in Eq.~\ref{eq:rho} that fix the density of the disk, $\rho_o$, the
base density (in $\rm g\,cm^{-3}$) and the power-law index $n$. For
this set of calculations, spectral types B0, B0.5, B1, B1.5, B2, and B3
were considered as these are the spectral types for which \citet{mar07a}
consider their SMC cluster samples complete.  Power-law indexes $n=2$,
$3$, and $4$ and 11 densities ranging from $10^{-13}$ to $10^{-10}$
$\rm g\,cm^{-3}$ were considered.

Two sets of Be star disk models were computed, one appropriate to the MW
and one appropriate to the SMC. Three basic changes distinguish between
the MW and SMC models:

[1] The abundance table used as input to {\sc bedisk} was changed to
reflect the composition of the gas. The adopted abundances for the
MW and SMC are given in Table~\ref{tab:abund}. The solar abundances
of \cite{asp05} were assumed for the Milky Way, and the abundance
scale for the SMC was taken mostly, although not exclusively, from
the VLT-FLAMES survey \citep{kor00,rol03,hun05,evan05,hun07,hun09}.
The adopted SMC abundances agree very well with the recent compilation
of \cite{evan09}. The general abundance pattern of the SMC is
$\Delta[Z/H]\approx -0.7$ \citep{mok07} but we keep the individual
abundances for each element.

[2] The input photoionizing radiation field was taken to be either
the \cite{Hubeny07} p00 for MW runs or m10 for the SMC runs. Test
calculations at the solar abundances of Table~\ref{tab:abund} show that
very similar temperatures result using either the p00 (solar) or m10
($1/10$ solar) photoionizing radiation fields. Hence, the small mismatch
between the overall $\Delta[Z/H]\approx -1.0$ abundances used to compute
the photoionizing radiation by \cite{Hubeny07} and actual SMC abundances
used here is likely of little consequence to the results.

\begin{figure}
\epsscale{1.0}
\plotone{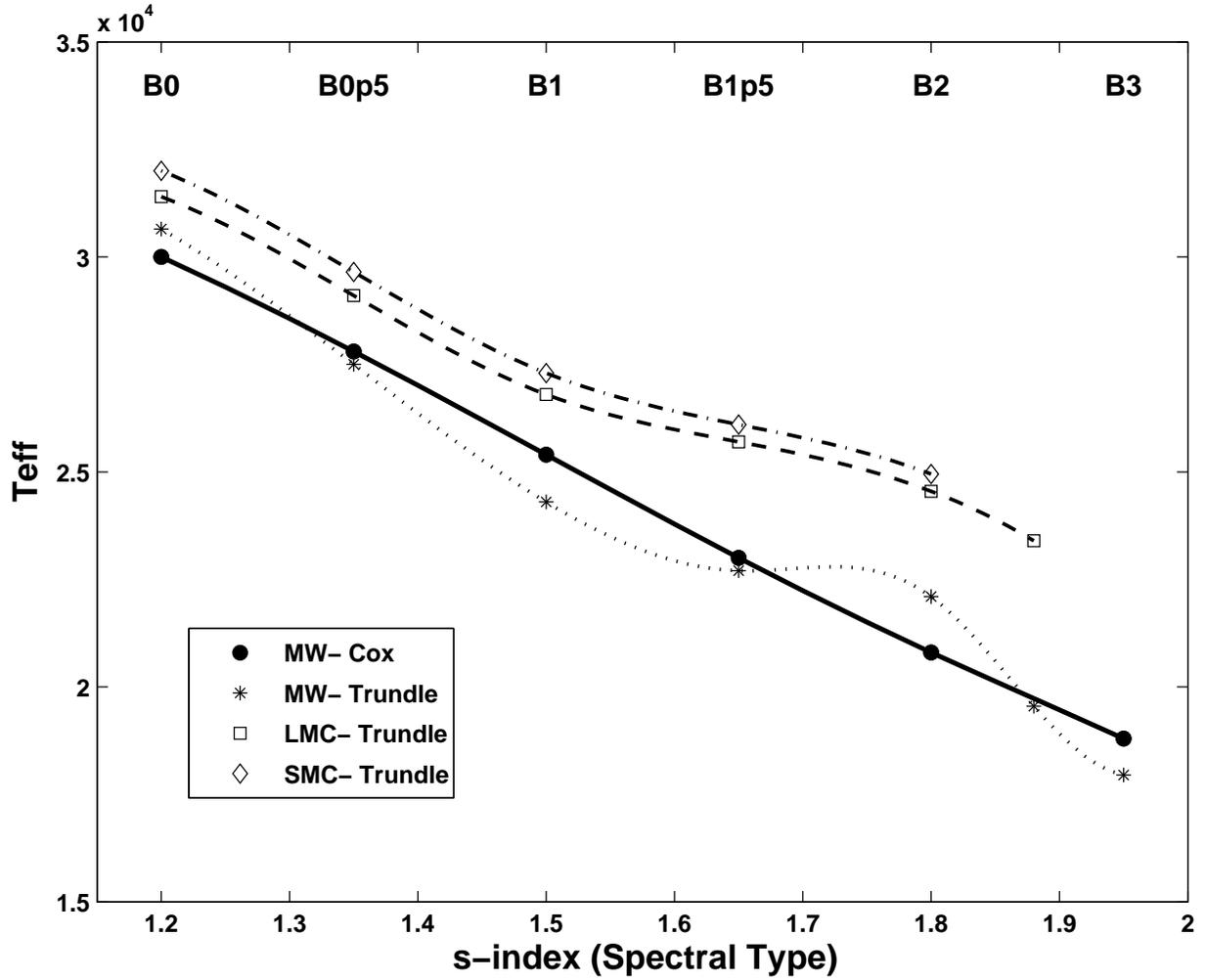}
%\plotone{teff_calibration.eps}
\caption{Comparison of the spectral type--$T_{\rm eff}$ calibration between the
MW and the SMC/LMC. Shown are the MW calibration of \cite{cox00}
and the LMC/SMC calibration of \cite{tru07}. The $x$-axis is the $s$-index of
\cite{cdj87}. The corresponding spectral types are indicated at the top
of the figure.\label{fig:teff}}
\vspace{0.1in}
\end{figure}

[3] It is well established that the spectral-type -- $T_{\rm eff}$
calibration differs between the MW and the SMC, with the SMC early-B
stars being hotter \citep{tru07}. This is shown in Figure~\ref{fig:teff}.
Here the Milky Way calibration of B dwarfs by \citet{tru07} agrees
well with the $T_{\rm eff}$ scale of \cite{cox00} (adopted in this
work) whereas the LMC and SMC calibrations are systematically hotter
by a few thousand degrees. Our adopted $T_{\rm eff}$ scales are given
in Table~\ref{tab:bstars}. Note that we do not adopt the \cite{tru07}
SMC $T_{\rm eff}$'s directly but increase our adopted MW $T_{\rm eff}$
calibration of \cite{cox00} by the amount found by \cite{tru07}. However,
the differences between our SMC $T_{\rm eff}$'s and that found directly
by \cite{tru07} are small.

Of these three differences between MW and SMC models, [1] and [3] have
the largest affect on the disk temperatures.

One caveat to our analysis is that we have ignored gravitational darkening
of the central B stars due to rotation. As a class, Be stars are well
established to be rapid rotators \citep{por96,yud01,por03}, although
how close to critical rotation (in which the equatorial velocity of
the star equals the orbital speed) is still contentious. \citet{tow04}
demonstrated that rotation rates above 80\% of critical rotation
(or $v_{\rm frac}\ge\,0.8$) may be difficult to detect because
gravitational darkening leads to a significant cooling of the star's
most rapidly rotating equatorial regions. Nevertheless, \citet{cra05}
used a statistical analysis {\it including the effects of gravitational
darkening\/} to demonstrate that the distribution of rotation speeds
for early Be stars (in the MW) is fairly uniform between 60 and 100\%
of critical. Hence for the MW, a representative value of 80\%, or $v_{\rm
frac}\approx 0.8$, is reasonable. \citet{mar07a} demonstrate that for the
SMC field NGC~330, Be stars rotate on average at $v_{\rm frac}=0.87$.
Recently, \citet{mcg11} have demonstrated that Be star disks become
systematically cooler as a result of gravitational darkening. Thus
any potential difference in mean rotation rates, $v_{\rm frac}$,
between the MW and SMC could result in a systematic {\it decrease\/}
in the temperature of SMC Be star disks. Fortunately, we can use
the results of \citet{mcg11} to estimate the expected effect. Taking
$v_{\rm frac}=0.8$ for the Milky Way and $v_{\rm frac}=0.9$ for the SMC
(rounding the observed value up for the maximal effect), Figures 4, 5
and 6 in \citet{mcg11} give a reduction in the density-weighted, global
disk temperature of $\approx\!300\,$K for an increase in $v_{\rm frac}$
from $0.8$ to $0.9$. As will be demonstrated in the subsequent sections,
this temperature decrease is much smaller (by at least a factor of five)
than the temperature {\it increase\/} predicted for SMC disks due to
effects [1] and [3] above. As a result, the neglect of gravitational
darkening is not a significant source of uncertainty.

\begin{deluxetable}{ccc}
\tablewidth{0pt}
\tablecaption{Adopted abundances for the MW and SMC.\label{tab:abund}}
\tablehead{
\colhead{Element} & \multicolumn{2}{c}{Abundance} \\
\colhead{~} & \colhead{MW}  & \colhead{SMC}
}
\startdata
He      & 10.92  & 10.9  \\
C       &  8.39  &  7.34 \\
N       &  7.78  &  6.97 \\
O       &  8.66  &  8.10 \\
Mg      &  7.53  &  6.76 \\
Si      &  7.51  &  6.77 \\
Fe      &  7.45  &  6.93 \\
\enddata
\tablecomments{The tabulated abundances are given as
$A=\log_{10}(N_x/N_{\rm H}) + 12$. The MW (solar) abundances are from \cite{asp05}. Sources for
the SMC are discussed in the text.}
\end{deluxetable}

%
% Fig 1
%

\section{Results: Temperature Structure}

\begin{figure}
\epsscale{1.0}
\plotone{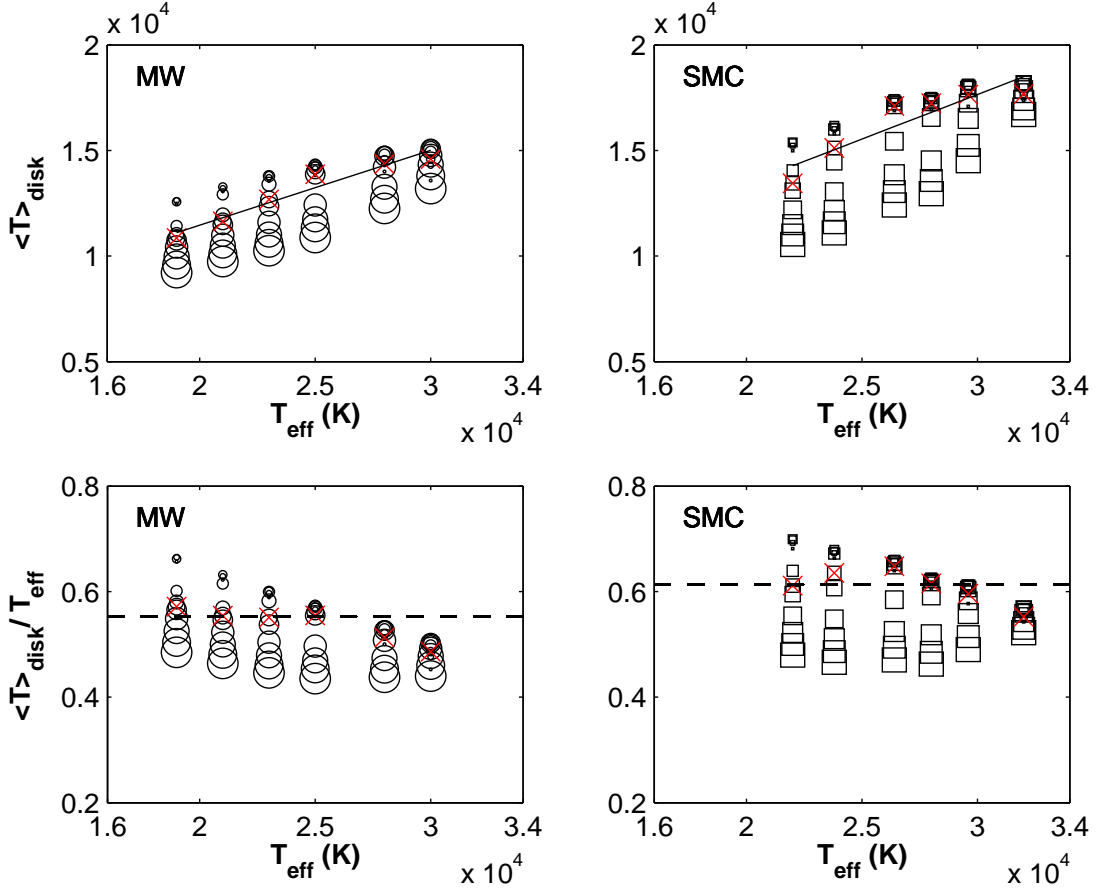}
%\plotone{allplots_n3.eps}
\caption{The density-weighted average disk temperature for the MW (left panels) 
and SMC (right panels). All models assumed $n=3$ in Eq.~\ref{eq:rho} 
and the symbol sizes indicate the size of $\rho_o$. The median at each 
spectral type is indicated by cross and the straight lines are fits to the medians.
\label{fig:allplots_n3}}
\vspace{0.1in}
\end{figure}

\begin{figure}
\epsscale{1.0}
\plotone{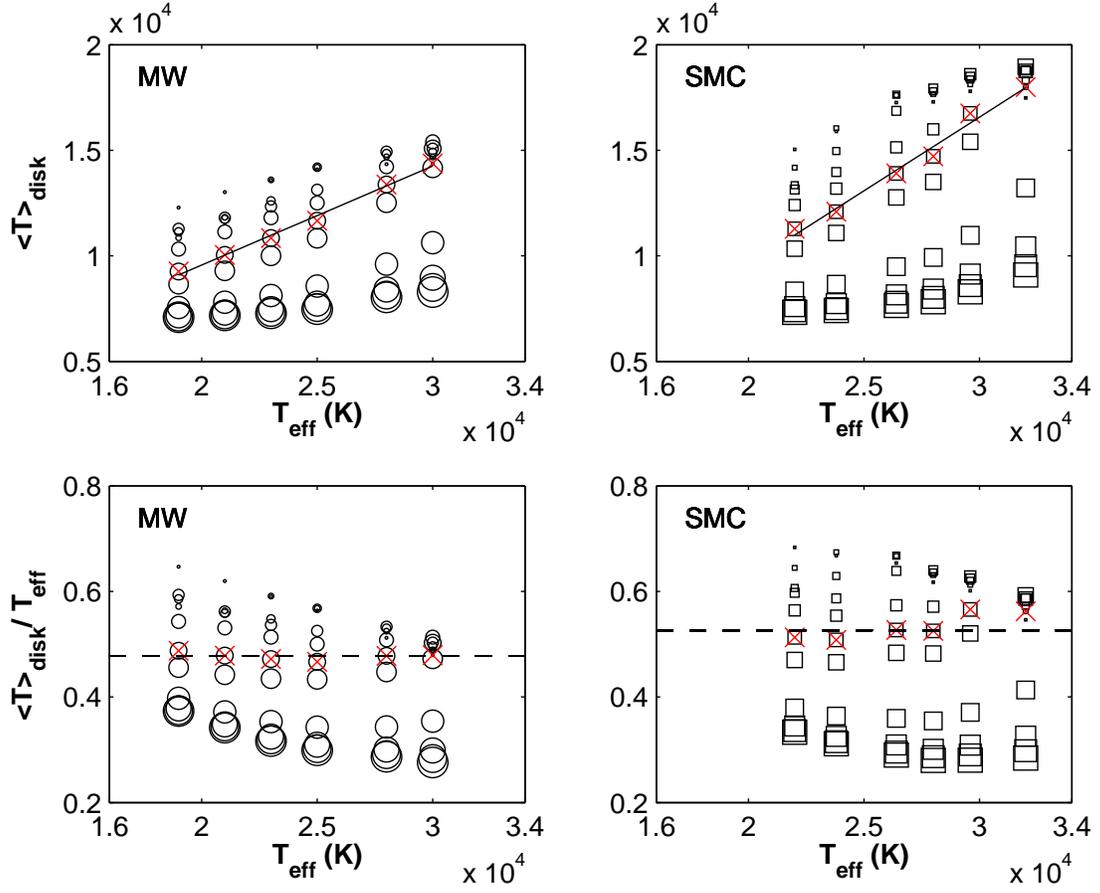}
%\plotone{allplots_n2.eps}
\caption{Same as Figure~\ref{fig:allplots_n3} except all models assumed $n=2$ in
Eq.~\ref{eq:rho}. Note the very low temperatures predicted by the highest 
density models. \label{fig:allplots_n2}}
\vspace{0.1in}
\end{figure}

In this section, we consider how disks of the same, fixed density
differ in their thermal structure due to the metallicity of the gas.
In order to compare a large number of models of differing spectral types
and density parameters, we will construct a single, global measure of the
disk temperature, namely the density-weighted, average disk temperature,
defined as 
\begin{equation} \label{T_mass_ave} <\!T\!>_{\rm disk} =
\frac{1}{M_{\rm{disk}}}\,\int_{\rm disk}\, T(R,Z)\,\rho(R,Z)\,dV \;.  
\end{equation}
Figure~\ref{fig:allplots_n3} shows this density-weighted disk temperature
as a function of the central B star's $T_{\rm eff}$ for both the MW and
SMC for all density models with a radial power-law index of $n\!=\!3$
in Eq.~\ref{eq:rho}.  Figure~\ref{fig:allplots_n2} shows a similar plot
but for the slower radial drop-off of $n\!=\!2$.  In both figures,
the global disk temperature is shown both as an absolute temperature
and as a fraction of the stellar $T_{\rm eff}$ (which differs between
the MW and SMC). For each $T_{\rm eff}$, temperatures and temperature
ratios are plotted for each of the 11 disk base densities, $\rho_o$,
considered. Immediately evident from the figures is the significantly
lower global disk temperatures for high values of $\rho_o$. Dense
disks develop a cool equatorial zone near the star \citep{mil98,sig09} and
this can significantly lower the density-weighted average disk temperature. This
effect is particularly noticeable in Figure~\ref{fig:allplots_n2} which
includes the densest disks considered in this work ($n=2$ and
$\rho_o=10^{-10}\;\rm g\,cm^{-3}$).

For $n=3$, the MW temperature ratios are reasonable well fit by
$<\!T\!>_{\rm disk}\!/T_{\rm eff}\approx 0.55$\footnote{\cite{sig09}
suggests a value of 0.6 as most appropriate for Be star disks. However,
the models in that work do not have a fixed vertical density structure
but self-consistently solve the equation of hydrostatic equilibrium. In
addition, they are of an odd type compared to the present work: pure
hydrogen/helium disks (i.e.\ zero metallicity) with a solar metallicity
(p00) photoionizing radiation field.} wheres the SMC disks are better fit
by $<\!T\!>_{\rm disk}\!/T_{\rm eff}\approx 0.61$. Thus the SMC disks
are even hotter than expected by the increased effective temperatures
of the SMC B stars.  This reflects the lower cooling rates in the SMC
disks due to the lower metallicity of the gas. Note that the densest
disks (the largest symbols) are not well-fit by the general median
trend and the fit is particularly poor for the densest disks with $n=2$
(Figure~\ref{fig:allplots_n2}); the cool equatorial zone significantly
lowers the density-weighted average temperatures.

\begin{figure}
\epsscale{1.0}
\plotone{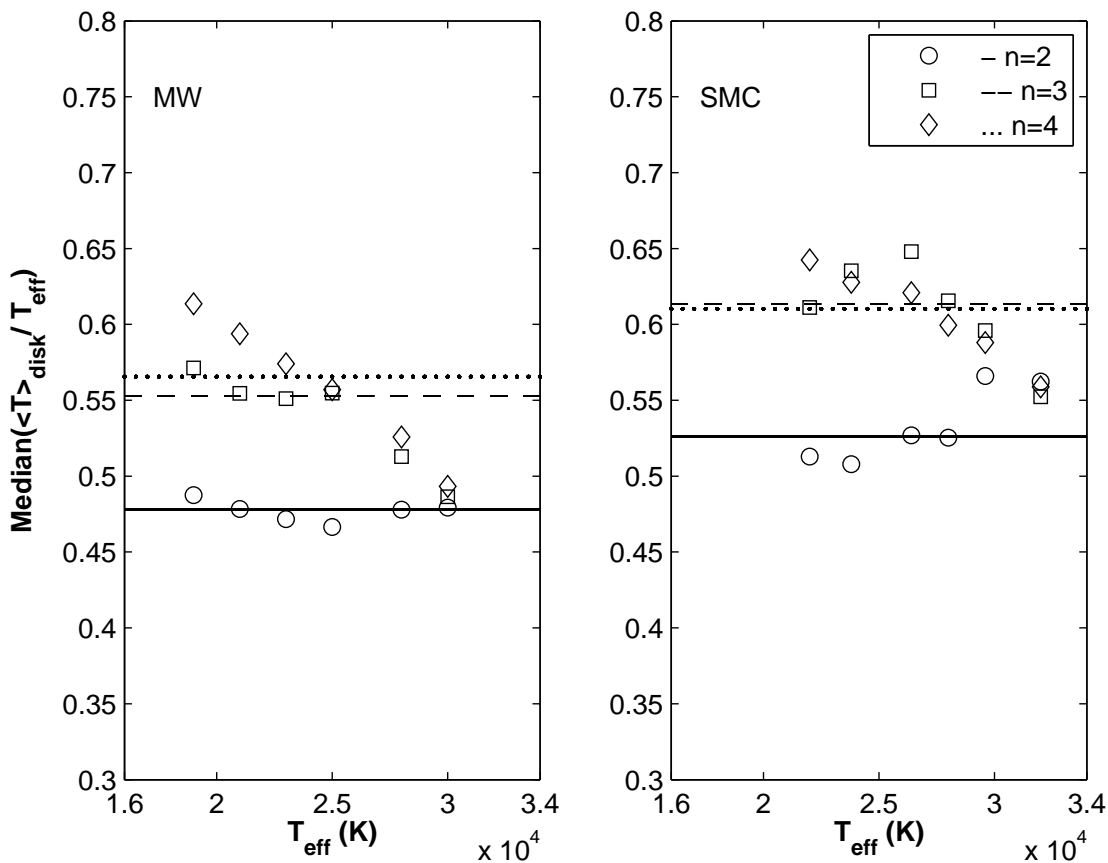}
%\plotone{medians.eps}
\caption{The median (over $\rho_o$) density-weighted disk temperatures
for the MW (left panel) and SMC (right panel) as a function of stellar
$T_{\rm eff}$. The symbols denote the power-law index in Eq.~\ref{eq:rho}
as $n=2$ (circles), $n=3$ (squares), and $n=4$ (diamonds).
The horizontal lines are the best constant fits to the medians.
\label{fig:median_ratio}}
\vspace{0.1in}
\end{figure}

This entire situation is summarized in Figure~\ref{fig:median_ratio}
which shows the disk temperature ratios, averaged over $\rho_o$, as a
function of $T_{\rm eff}$ for the three power-law indexes considered,
$n=2$, $3$, and $4$.  There is little difference between the $n=3$
and $4$ models whereas the $n=2$ models are significantly cooler due to
their larger densities.  Note the median ratio over all $T_{\rm eff}$
considered is always higher in the SMC. The numerical values for the
median temperature ratios are given in Table~\ref{tab:ratios}.

%
% Fig 2
%

%
% Fig 3
%

\begin{figure}
\epsscale{1.0}
\plotone{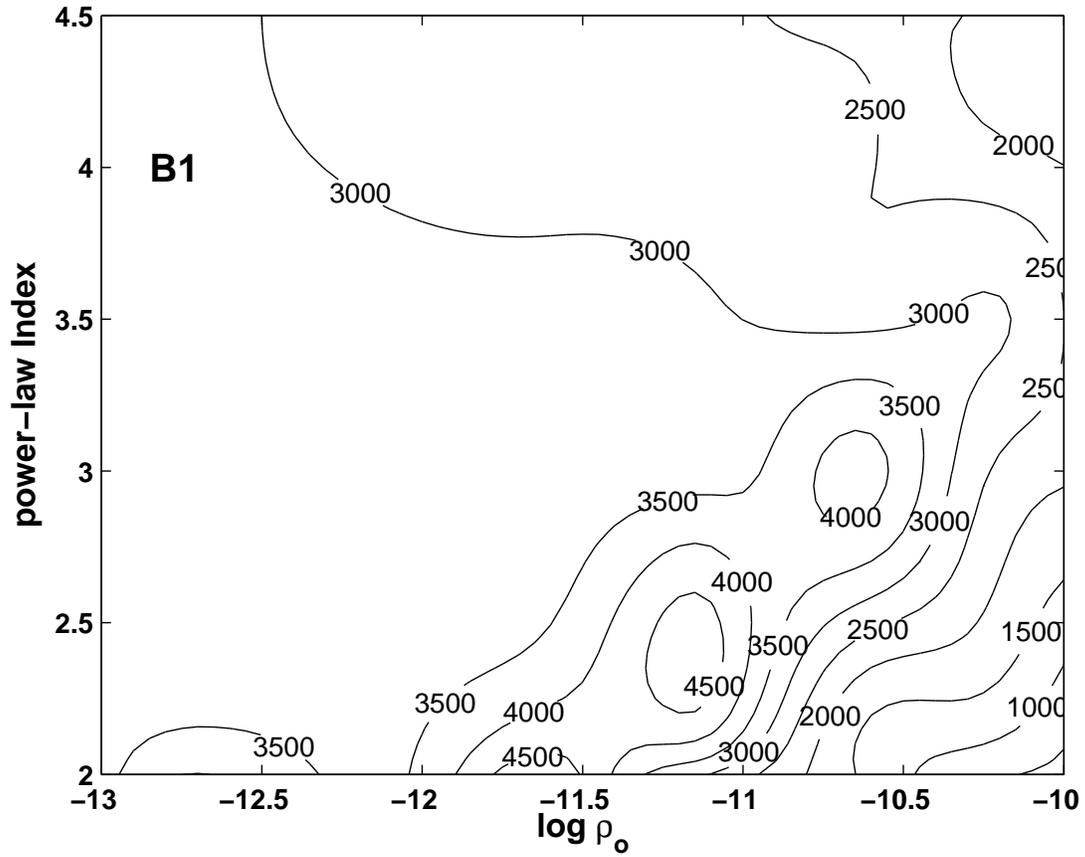}
%\plotone{npwr_VS_Rho0_tm_smc-tm_p00_B1.eps}
\caption{Difference (SMC-MW) in the density-weighted average disk temperature, 
Eq.~\ref{T_mass_ave},
as a function of the disk base density, $\rho_o$, and power-law index, $n$,
for spectral type B1. The contours are of equal temperature difference in degrees K.
\label{fig:B1_tdiff}}
\vspace{0.1in}
\end{figure}

Figure~\ref{fig:B1_tdiff} shows as a contour plot of the difference
between SMC and MW density-weighted average disk temperature as a function
of the disk density parameters $\rho_o$ and $n$ for the spectral type
B1. Differences of up to $4500\,$K are seen (SMC hotter) with the largest
values occurring for intermediate disk densities $\rho_o$. The
location of the maximum temperature difference in this plot tends to
move to higher $\rho_o$ for larger $n$. The temperature difference rapidly drops
for the densest models considered ($\log\rho_o>-10.4$ and $n<2.5$) as the
density-weighted temperature in this region is dominated by the very cool
equatorial zone close to the star; this occurs in the most optically thick
portion of the disk where the sensitivity to the gas metallicity is least.

\begin{figure}
\epsscale{1.0}
\plotone{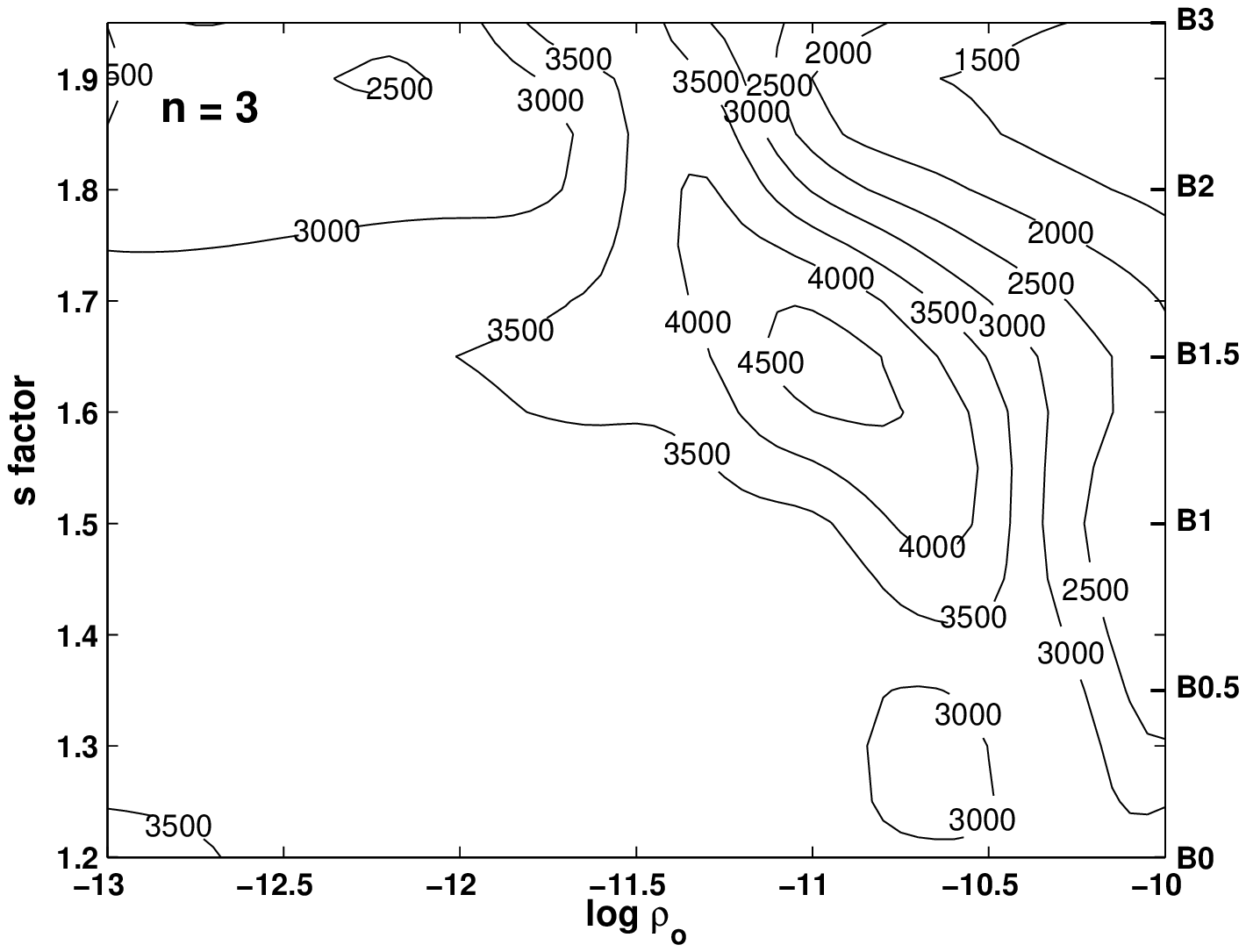}
%\plotone{s_logrho0_n3.eps}
\caption{Difference (SMC-MW) in the density-weighted average disk temperature
as a function of the disk
base density, $\rho_o$, and the spectral type of the star, s-factor, for power-law
index $n=3$. The contours are of equal temperature difference in degrees K.
\label{fig:sfac_tdiff}}
\vspace{0.1in}
\end{figure}

Figure~\ref{fig:sfac_tdiff} again shows the temperature difference between
the SMC and MW models but this time as a function of spectral type and
$\rho_o$ for a fixed power-law index of $n=3$. The spectral type is
plotted as the $s$-factor of \cite{cdj87} (namely 1.20 for B0, 1.35 for
B0.5, 1.50 for B1, 1.65 for B1.5, 1.80 for B2 and 1.95 for B3). Again the
largest temperature differences are for the intermediate density models
with the largest absolute differences occurring near spectral type B1.5,
near the peak of the Be star distribution with spectral type.

%
% Fig 4
%

\begin{deluxetable}{lrr}
\tablewidth{0pt}
\tablecaption{Disk temperatures ratios averaged over all 
spectral types, $\rho_o$, and $R_d$, as a function of power-law index $n$.
The quoted dispersion is $1\,\sigma$.
\label{tab:ratios}}
\tablehead{
\colhead{Index $n$} & \multicolumn{2}{c}{\hrulefill~$<\!T\!>_{\rm disk}/T_{\rm eff}$~\hrulefill} \\
\colhead{~}         & MW     & SMC
}
\startdata
4.0  & $0.57\pm0.07$  & $0.61\pm0.08$ \\
3.0  & $0.55\pm0.06$  & $0.61\pm0.08$ \\
2.0  & $0.48\pm0.10$  & $0.53\pm0.13$ \\
\enddata
\end{deluxetable}

%
% Fig 5
%

%
% Fig 6
%

\section{Results: H$\alpha$ Equivalent Width}

While interesting, the previous results are of limited practical use
because the thermal structure of a circumstellar disk is not directly
observed. The main observable for Be stars is the equivalent width
(in emission) of the H$\alpha$ line ($\lambda\,6562.8\,$\AA) in the
spectrum. Hence it is of direct practical interest to compare the
H$\alpha$ emission line equivalent widths between the MW and SMC models.
All current methods designed to find and measure the Be star fraction in a
stellar cluster implement some technique to select candidate emission-line
stars based on a measurement of H$\alpha$.  For example, \cite{mar10}
use very low-resolution slitless spectroscopy and estimate that their
survey of Be stars in SMC clusters detects Be stars with H$\alpha$
emission equivalent widths in excess of $10\,$\AA\ or peak intensities
more than twice the adjacent continuum.

To this end, we have used the {\sc beray} code of \cite{sig10} to compute
the H$\alpha$ line profile and equivalent width (in \AA) for each of
the MW and SMC disk models of the previous section. The {\sc beray}
code solves the equation of radiative transfer along a series of rays
through the star-disk system to compute both resolved monochromatic
images and unresolved spectra.  The H$\alpha$ calculations used the
$n\!=\!2$ and $3$ level populations of hydrogen from the {\sc bedisk}
thermal solution. Rays that terminate on the stellar surface use the
Doppler shifted photospheric LTE H$\alpha$ line profiles\footnote{These
profiles were computed in stellar atmospheres assuming the solar
abundance. While the profile corresponding to the required stellar
temperature ($T^{\rm MW}_{\rm eff}$ or $T^{\rm SMC}_{\rm eff}$) was used,
this still represents a small inconsistency.  We suspect that this effect
is smaller than the uncertainty introduced by the use of LTE, as opposed
to non-LTE, profiles. Nevertheless as we are really interested only in
models with strong emission in H$\alpha$ for which the disk dominates,
the adopted photospheric profile is not a major source of uncertainty.}
of \cite{bar03} for the initial boundary condition, while rays that pass
entirely through the disk assume no incident radiation.  The formal
solution to the transfer equation was performed using the H$\alpha$
source function and the bound-free and free-free continuous opacity and
emissivity sources at the wavelength of H$\alpha$.  Electron scattering
was also included using the mean intensity at each position in the disk as
the source function for the scattering emissivity. The disk was assumed to
be in pure Keplerian rotation and the frequency-angle dependent transfer
equation was solved in the observer's frame \citep{mil78}.

Calculating the H$\alpha$ line profile adds two parameters to the
problem, the inclination of the Be system to the line of sight ($i=0^o$
indicates a pole-on star/disk, and $i=90^o$, an edge-on one) and the
outer radius of the disk, $R_{d}$ (quoted here in units of the stellar
radius).\footnote{The thermal solution for the disk structure, computed
by {\sc bedisk}, assumed $R_d=30\,R_*$. In {\sc beray}, smaller choices
for $R_d$ were enforced by multiplying all level populations beyond $R_d$
by $10^{-5}$. The inner radius of all disks was taken to be the stellar
radius.} Values for these parameters were taken to be $i=15$, $30$,
$45$, $60$, and $75^o$, and $R_d=5$, $10$, $15$, and $20\,R_*$. For
each spectral type, there were a total of 660 models representing all
permutations of the parameters $\rho_o$, $n$, $R_d$ and $i$. 
%However, not all of these models had disks that produced detectable emission
%in H$\alpha$.

\begin{figure}
\epsscale{1.0}
\plotone{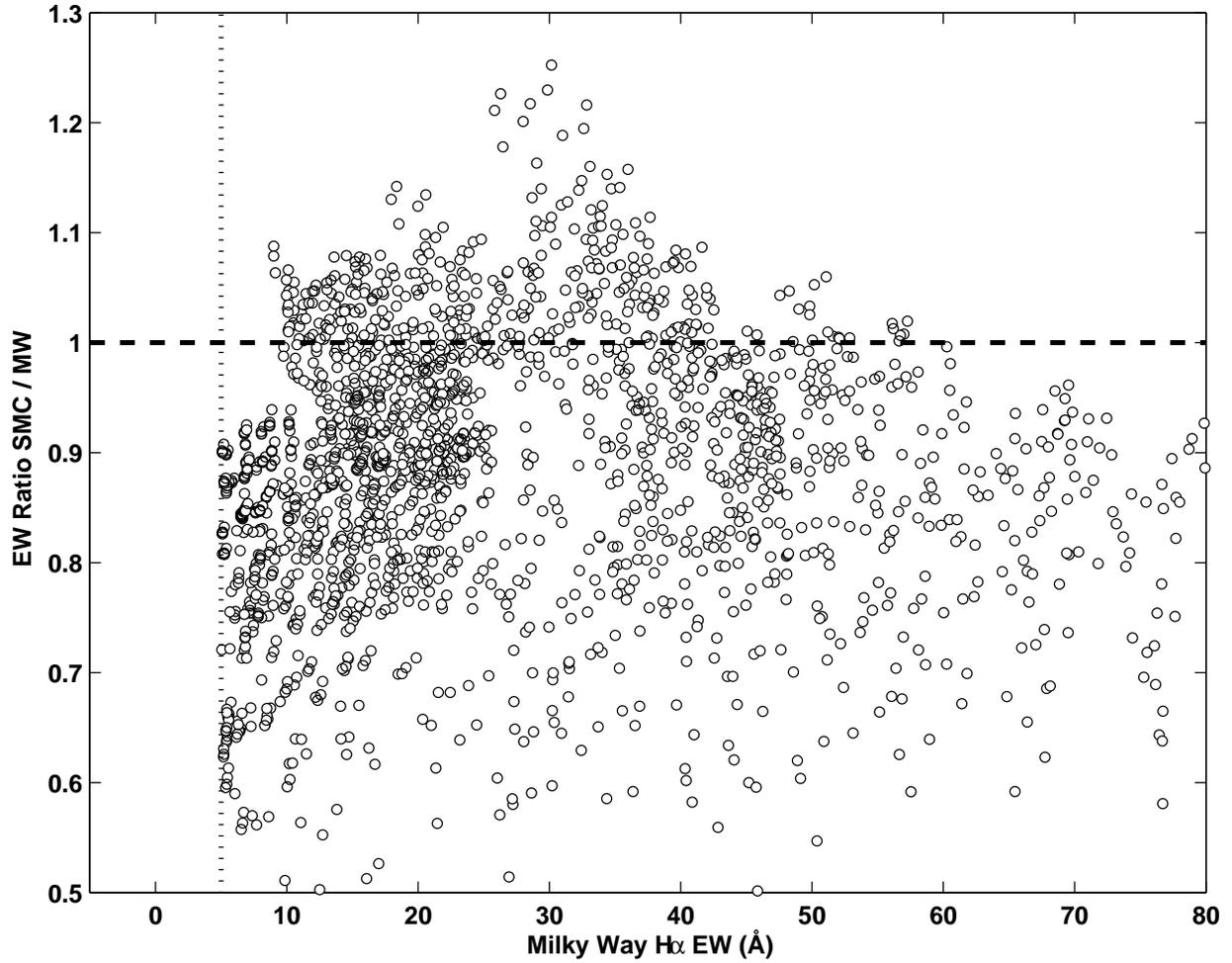}
%\plotone{eqw_ratio_new.eps}
\caption{Ratio of the SMC to MW H$\alpha$ equivalent width as a 
function of the MW equivalent width. Only models predicting
a MW equivalent width of $>5\,$\AA\ are plotted (see text).\label{fig:eqw_ratio}}
\vspace{0.1in}
\end{figure}

The ratio of the SMC and MW H$\alpha$ equivalent widths is shown in
Figure~\ref{fig:eqw_ratio}. The trend is clear: H$\alpha$ is generally
{\it weaker\/} in the SMC compared to a disk of the same density structure
in the MW. Over all models satisfying $\rm EW(H\alpha)>5\,$\AA, only 366
out of 2047 models (or 18\%) had a stronger H$\alpha$ line in the SMC;
the median SMC to MW equivalent width ratio is $0.89$.

For early-type Be stars, disk temperatures usually exceed $10^4\,$K
and increased temperatures tend to reduce the population of the excited
states of hydrogen due to increased ionization. This results in a weaker
H$\alpha$ line. Nevertheless, the temperature structure of the denser
Be disk models is quite complex \cite[see, for example,][]{sig09} and
some of these models can produce a stronger H$\alpha$ line in the SMC.
Table~\ref{tab:fraction_sp} gives the fraction of the models in which
the SMC H$\alpha$ equivalent width was larger than the MW model (for an
identical disk) as a function of spectral type. This fraction increases
strongly for later spectral types (more than doubling over the range
considered) showing that the cooler disks around later spectral types
have a much larger fraction of increased H$\alpha$ strengths in the SMC.

Table~\ref{tab:fraction_i} shows an additional interesting result.
Over all models (spectral type and density structure), there is a very
strong dependence of the fraction of models that predict a stronger
H$\alpha$ EW in the SMC on the viewing angle. Increased emission is
essentially eliminated (fraction less than 10\%) for models seen more
``disk-on" with $i\ge60^o$. Note that the maximum viewing inclination in
this work, $75^o$, does not produce obvious shell-spectra characteristic
of Be star disks seen nearly edge-on.\footnote{Be shell stars are not
considered in the present work. The fraction of Be shell
stars in the MW is $\approx\,23$\% \citep{han96}. The fraction in the SMC
is estimated to be $\approx\,16$\% \citep{mar07b}.}

\begin{deluxetable}{lccc}
\tablewidth{0pt}
\tablecaption{Number and fraction of Be star models with a larger H$\alpha$ equivalent
width in the SMC as a function of the spectral type of the central B star.\label{tab:fraction_sp}}
\tablehead{
\colhead{Spectral Type} & \colhead{H$\alpha$ Larger} & \colhead{H$\alpha$ Smaller} &
\colhead{Larger Fraction}
%\colhead{~} & \colhead{in SMC} & \colhead{in SMC} & \colhead{~}
}
\startdata
B0.5 &  38 & 325 & 10\% \\
B1   &  57 & 346 & 14\% \\
B1.5 &  82 & 341 & 19\% \\
B2   &  81 & 346 & 19\% \\
B3   & 108 & 323 & 25\% \\
\enddata
\tablecomments{Only models predicting $\rm EW(H\alpha)>5\,$\AA\ emission are included.}
\end{deluxetable}

\begin{deluxetable}{lccc}
\tablewidth{0pt}
\tablecaption{Number and fraction of Be star models with a larger H$\alpha$ equivalent
width in the SMC as a function of the viewing angle $i$.\label{tab:fraction_i}}
\tablehead{
\colhead{Inclination} & \colhead{H$\alpha$ Larger} & \colhead{H$\alpha$ Smaller} &
\colhead{Larger Fraction}
}
\startdata
$15^o$ & 133 & 332 & 29\% \\
$30^o$ & 110 & 316 & 26\% \\
$45^o$ &  80 & 347 & 19\% \\
$60^o$ &  33 & 345 &  9\% \\
$75^o$ &  10 & 341 &  3\% \\
\enddata
\tablecomments{Only models predicting $\rm EW(H\alpha)>5\,$\AA\ emission are included.}
\end{deluxetable}

\begin{figure}
\epsscale{1.0}
\plotone{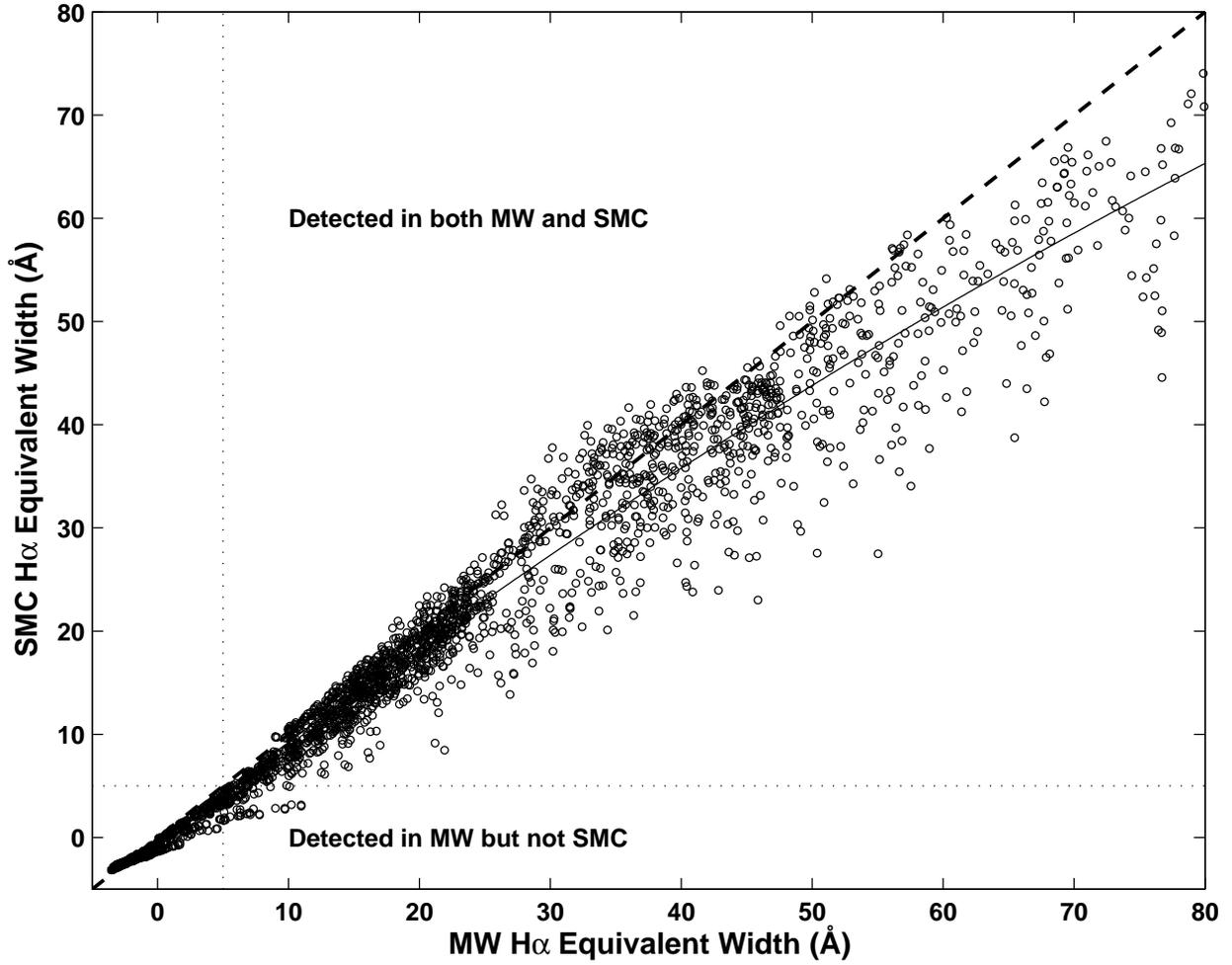}
%\plotone{smc_mw_eqw_new.eps}
\caption{The SMC H$\alpha$ equivalent as a function of the MW H$\alpha$ 
equivalent width. Detection
limits of $5\,$\AA\ are as indicated, as are regions where disks would be
detected {\it both\/} in the MW and SMC and {\it only\/} in the MW.
The dashed line is of unit slope and the solid line is a quadratic,
least-squares fit to the
calculations.\label{fig:eqw_line}}
\vspace{0.1in}
\end{figure}

The difference in H$\alpha$ equivalent widths between the MW and SMC
could potentially affect the comparison of the fraction of Be stars in
clusters. Even if both the MW and the SMC had the same distribution
of disk density parameters (i.e.\ $\rho_o$, $n$ and $R_d$), there
would nevertheless be a different distribution of H$\alpha$ line
strengths because of the different temperature structure of the disks.
Figure~\ref{fig:eqw_line} shows the SMC H$\alpha$ equivalent width
plotted against the MW value for a disk of the same density structure. If
one assumes a selection method for candidate Be stars with the same
threshold in the MW and SMC, there will be a population of stars with
weaker emission in the SMC that fall below the detection limit whereas
the MW counterparts  would be counted.  This effect is quantified in
Table~\ref{tab:be_miss} which gives the missed percentage of Be stars
among the SMC models as a function of H$\alpha$ detection thresholds
of between $2$ and $15\,$\AA. As can be seen, this effect is small with
the missed percentage rising to $\approx\!10$\% only for thresholds
as  large as $15\,$\AA. For thresholds between 5 and $10\,$\AA, the
prediction is about 5\%.  Hence despite the rather large temperature
differences between the SMC and MW disks, the systematic difference in
the H$\alpha$ equivalent widths is not a serious source of bias in Be
star candidate counts.

\begin{deluxetable}{cccc}
\tablewidth{0pt}
\tablecaption{Percentage of Be stars missed in the SMC as a function of H$\alpha$
equivalent width threshold.\label{tab:be_miss}}
\tablehead{
\colhead{Threshold EW} & \multicolumn{2}{c}{Number of Detected Be Stars} & \colhead{Percent Missed}\\
\colhead{(\AA)} & \colhead{MW}  & \colhead{SMC} & \colhead{~}
}
\startdata
 2.0 & 2210 & 2157 & 2.4\% \\
 5.0 & 2047 & 1939 & 5.2\% \\
10.0 & 1774 & 1670 & 5.9\% \\
15.0 & 1458 & 1324 & 9.2\% \\
\enddata
\end{deluxetable}

Finally, we compare the H$\alpha$ equivalent width distributions predicted
by our models of both the MW and SMC with the observational results of
\citet{mar07b}. In this comparison, we include all models producing
emission (i.e.\/ all values of $n$, $\rho_o$, $R_d$ and $i$ which
result in $\rm EW(H\alpha)>0$) except those with the smallest disks,
$R_d=5\,R_*$.  In constructing the histograms, we have weighted each
model based on its inclination as follows: for random inclinations, the
probability of observing an inclination between $i$ and $i+di$ is $\sin
i\,di$. Therefore, we have assumed that each of the five values of $i$
considered is the centre of a (non-overlapping) bin and have weighted
each model by the fractional area in that bin.\footnote{The bins are
(in degrees) $[0,22.5)$, $[22.5,37.5)$, $[37.5,52.5)$, $[52.5,67.5)$,
and $[67.5,90]$ which have fractional areas $0.0761$, $0.1305$, $0.1846$,
$0.2261$, and $.3827$, respectively. As an example, for each 10 stars
in the first bin, there should be 50 stars in the last bin.}

\begin{figure}
\epsscale{1.0}
\plotone{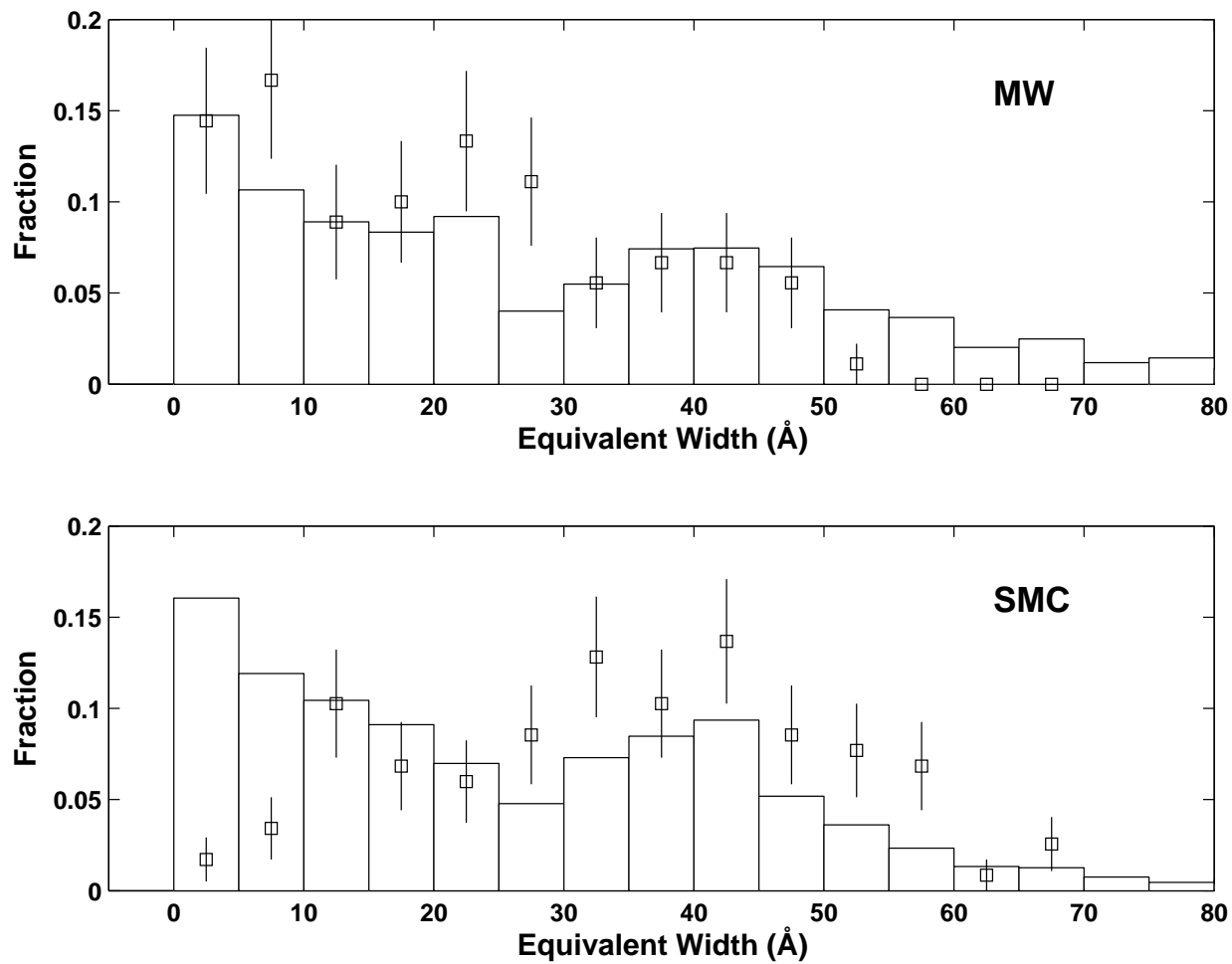}
%\plotone{hist_obs_full.eps}
\caption{Histograms of the H$\alpha$ equivalent widths predicted by the MW and
SMC models. All models predicting emission ($\rm EW>0$) are included with the
exception of the models with very small disks ($R_d=5\,R_*$). The fraction represented
by each EW bin is relative to the total number of models satisfying $\rm EW>0$.
The squares are
the observations of \cite{mar07b} (in the same EW bins) with $\sqrt{N}$ error bars added.
\label{fig:eqw_hist}}
\vspace{0.1in}
\end{figure}

\begin{figure}
\epsscale{1.0}
\plotone{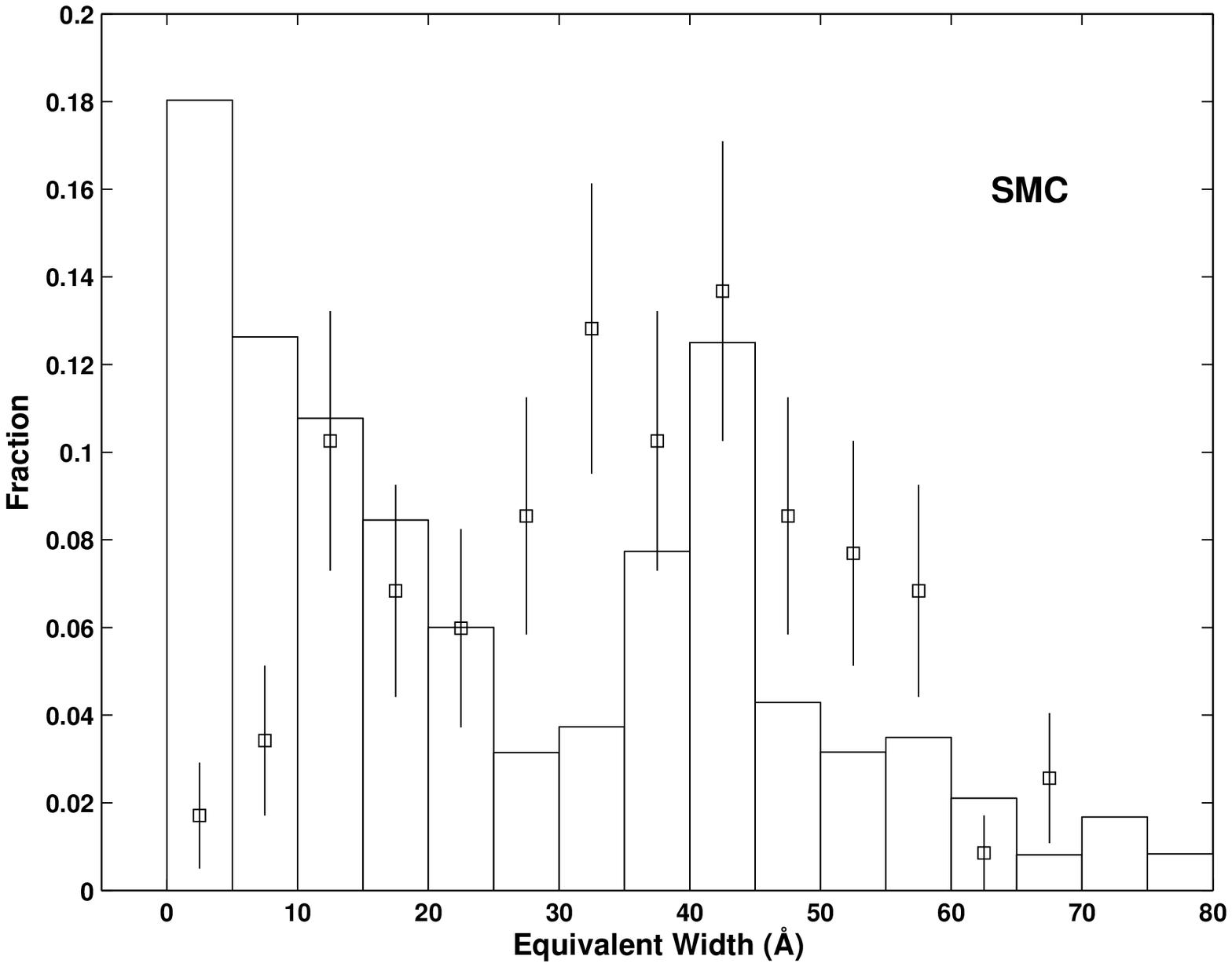}
%\plotone{hist_rd20_smc.eps}
\caption{A Histogram of the H$\alpha$ equivalent widths predicted by the
SMC models with $R_d=20\,R_*$. 
All models predicting emission ($\rm EW>0$) 
are included and the fraction represented by each bin is relative to the total.
The squares are
the observations of \cite{mar07b} (in the same EW bins) with $\sqrt{N}$ error bars added.
\label{fig:eqw_hist_rd20}}
\vspace{0.1in}
\end{figure}

The results are shown in Figure~\ref{fig:eqw_hist}. For
the Milky Way, \citet{mar07b} adopt the equivalent width measurements
of \citet{and82}, \citet{and83}, and \citet{dac92}. As noted, the only
``fine-tuning" of our models is the exclusion of the small $R_d=5\,R_*$
disks; these disks produce a large peak for $\rm EW(H\alpha)<10\,$\AA\
which is not seen in the observations. Such small disks, when they seem to
occur in nature, are usually attributed to binary truncation \citep[for
example see the case of $\alpha\;$Ara as discussed by][]{mei09}. Given
this, the predicted MW equivalent width distribution gives a reasonable
match to the observations.  However, we do note that the sample of Milky
Way observations finds few widths above $55\;$\AA\ whereas such
values are represented in our models. For the SMC, the models produce
a poorer match to the observations of \citet{mar07b}. We suspect the
large mismatch for $<10\;$\AA\ may be incompleteness in the observations.
However, the models under-predict the fraction of large equivalent widths
($>25\;$\AA) by a significant amount and this difference seems to be
a difference between the MW and SMC, as noted by \citet{mar07b}. As
suggested by these authors, this and other observational evidence
suggests that Be star disks are systematically larger in the SMC. Indeed,
Figure~\ref{fig:eqw_hist_rd20} shows that we can produced a peak in the
SMC distribution near $40\;$\AA\ by restricting the model set to only the
largest disks considered, $R_d=20\,R_*$, although the peak is not wide
enough. We suspect that including even larger disks might improve the fit.

%
% Fig 7
%

%
% Fig 8
%

%
% Fig 9
%

%
% Fig 10
%

%\section{Results: Hydrostatic Models}

\section{Conclusions}

We have computed a large set of Be star disk models appropriate to the
average metallicities of the Milky Way and the SMC. We have shown that for
a disk of identical density structure, SMC disks are systematically hotter
than MW disks, typically by several thousand degrees. This difference
is attributable to the higher $T_{\rm eff}$ scale of massive stars in
the SMC and the lower metallicity of the SMC disk gas, which reduces its
ability to cool. For all of the considered models, the H$\alpha$ emission
equivalent width is generally smaller in the SMC as compared to MW disks
of identical density. However, this systematic difference is not predicted
to affect comparisons of Be star fractions between the MW and SMC.

We also show that the H$\alpha$ equivalent width distributions of our
models is in reasonable agreement with the known MW distribution.
However, a similar distribution of underlying parameters does not
seem to fit the SMC. We confirm the suggestion of \citet{mar07b}
that systematically larger Be stars disks in the SMC may explain this
discrepancy. It would be desirable to put this interesting result on a
sounder footing by directly determining the disk parameters for a sample
of SMC Be star disks via the analysis of high-resolution spectra. Such a
study would directly give the distribution of disk parameters ($\rho_o$,
$n$ and $R_d$) which could be compared to a similar sample in the MW
and would allow a more detailed investigation of how lower metallicity
affects these circumstellar disks.

\acknowledgments
We would like to thank the referee for suggesting many improvements to
this paper.  This work is supported by the Canadian Natural Sciences and
Engineering Research Council (NSERC) through a Discovery Grant to TAAS.

\end{document}